\documentclass[12pt,preprint]{aastex}

\slugcomment{} \shorttitle{Influence of} \shortauthors{Ma, Yuan, &
Wang}

\begin{document}

\title{Influence of the Magnetic Coupling Process on the Advection
Dominated Accretion Flows around Black Holes}

\author{Ren-Yi Ma, Feng Yuan}
 \affil{Shanghai Astronomical Observatory, Chinese Academy
of Sciences, 80 Nandan Road, Shanghai 200030, China;
Joint Institute for Galaxy and Cosmology (JOINGC) of SHAO and USTC;
Email: ryma@shao.ac.cn; fyuan@shao.ac.cn}
 \and
 \author{ Ding-Xiong Wang}
  \affil{Department of Physics, Huazhong
University of Science and Technology,Wuhan, 430074, China;
 Email: dxwang@hust.edu.cn}

\begin{abstract}

A large-scale closed magnetic field can transfer angular momentum
and energy between a black hole (BH) and its surrounding accretion
flow. We investigate the effects of this magnetic coupling (MC)
process on the dynamics of a hot accretion flow (e.g., an advection
dominated accretion flow, hereafter ADAF). The energy and angular
momentum fluxes transported by the magnetic field are derived by an
equivalent circuit approach. For a rapidly rotating BH, it is found
that the radial velocity and the electron temperature of the
accretion flow decrease, whereas the ion temperature and the surface
density increase. The significance of the MC effects depends on the
value of the viscous parameter $\alpha$. The effects are obvious for
$\alpha=0.3$ but nearly ignorable for $\alpha=0.1$. For a BH with
specific angular momentum, $a_*=0.9$, and $\alpha=0.3$, we find that
for reasonable parameters the radiative efficiency of a hot
accretion flow can be increased by $\sim 30\%$.

\end{abstract}


\section{Introduction}

As a variant of Blandford-Znajek (BZ) process \citep{BZ77}, the
magnetic coupling (MC) between the central rotating black hole (BH)
and its surrounding accretion disks has received much attention
\citep[e.g.][]{Blandford99, Li00,Li02,Wang02}. By virtue of the
large-scale closed magnetic field lines that connect the BH and its
surrounding disk, the MC process conveys energy and angular momentum
between the BH and the disk. \citet{Li02} showed that in the
standard disk \citep[SSD,][]{SS73,NT73,PT74} case the MC process may
change the local radiative flux significantly.

Apart from SSD, ADAF is another important model of the accretion
flow (\citealt{NY94,NY95,A95}; see \citealt*{NMQ98} and
\citealt*{KFM98} for reviews). It has been applied to a number of
accreting BH systems and successfully explains their spectral
characteristics
(see Narayan 2005 and Yuan 2007 for recent reviews). In ADAF models
the thickness of the accretion flow is of the same order as radius,
i.e. $H \sim r$, which means the large-scale field is easier to form
in an ADAF than in a SSD \citep{Tout96,Livio99}. So it is
interesting to investigate the influences of the MC process on an
ADAF. Very recently, \citet{Ye07} discussed this problem based on
the self-similar solution of the ADAF. In this paper, we investigate
the influences of the MC process on ADAFs through global solutions.

To properly assess the dynamical effects of large-scale magnetic
fields on the accretion flow, one needs to obtain the fluids and the
fields at the same time by solving the transfield equation, which is
a nontrivial nonlinear partial differential equation with singular
surfaces and free functions \citep{U04,U05}. An alternative way is
MHD simulations
\citep[e.g.][]{Hawley00,HB02,Koide03,DeV03,MG04,Hirose04}. However,
both of these approaches are complicated. \citet{Lai98} and
\citet{Lee99a,Lee99b} adopted a phenomenological approach to
research the magnetic coupling between the neutron star and its
surrounding slim disk. They specified an ansatz for the magnetic
fields and then numerically solve the basic equations of the
accretion flow. In their model the disk is geometrically thin,
$\partial /
\partial r\sim 1 / r \ll 1 / H\sim
\partial / \partial z$, and so the expressions of electromagnetic
forces can be much reduced by omitting $\partial / \partial r$
terms. However, an MCADAF is thick and the
expression of the electromagnetic force is complex. Here for simplicity
we treat the MC process as a source of energy and angular
moment without considering the radial and vertical components of the
electromagnetic force in the momentum equations.

We derive the energy and angular momentum fluxes in the Kerr metric
by using the approach of equivalent circuit \citep{MT82}. But for
simplicity a pseudo-Newtonian potential of a rotating black hole
given by \citet{Muk02} is adopted when we solve the solutions of the
accretion flow.

In Section 2, we describe the MCADAF model and calculate the energy
and angular momentum fluxes transferred by the magnetic field. In
Section 3, we write down the basic equations describing the MCADAF.
The numerical results are presented in Section 4 and Section 5 is
devoted to a summary and discussions. Throughout this paper the
geometric units c=G=1 are used.

\section{MCADAF Model}

We assume the ADAF is stationary and axisymmetric. The ADAF extends
from the outer edge, $r_{out} $, to the BH horizon $r_H$. There are
two kinds of magnetic fields in this model, i.e., large-scale closed
magnetic field that connects the BH with the ADAF and small-scale
tangled magnetic field, with the former contributing to the MC
process and the latter to the viscosity. We assume these two kinds
of fields work independently. If not mentioned we refer the magnetic
field as the large-scale closed one hereafter.  The region between
the BH and the ADAF is assumed to be ideally conducting and
force-free.

The field lines are supposed to distribute in the ranges of $(r_H
,r_{out})$ on the disk and $(0,\theta _0 )$ on the horizon. Due to
the lack of knowledge about the magnetic field around the BH, we
assume that the field threading the BH is constant, i.e. $B_H
(\theta ) = const$. The field threading the ADAF is assumed to
decrease with $r$ following a power law form, but within the
marginally stable orbit, $\la r_{ms}$, the radial velocity of the
accretion flow increases much faster thus the filed is likely to
increase with radius. Given this consideration, we assume the field
has the following distribution,

\begin{equation}
\label{eq1} B_z (r) = B_0 F(r) = \left\{ {{\begin{array}{l@{\quad
\mbox{for} \quad}c}
 B_0 \exp ( {r / r_p - 1}) & r_H < r \le r_p
 \\

 B_0 ( {r / r_p } )^{ - n} & r_p < r \le r_{out}
\end{array} }} \right.
\end{equation}
Here $r_H=M\left( {1 + \sqrt {1 - a_\ast ^2 } } \right)$ denotes the
radius of the BH horizon, $a_* $ is the dimensionless spin parameter
of the BH, and $r_p=r_H+\lambda(r_{ms}-r_H)$.

 \citet{MSL} gave an
estimation of $B_H $ with the balance between the ram pressure of
the falling material and the magnetic pressure, i.e., $B_H^2 / 8\pi
\sim \rho \sim \dot {M}_D / \left( {4\pi r_H^2 } \right)$. Since the
ram pressure can be larger than the magnetic press, we introduce a
parameter $c_B $ to indicate the strength of the magnetic field
threading the horizon as

\begin{equation}
\label{eq2}
 B_H = c_B \sqrt {2\dot {M}} / r_H ,\quad 0 \le c_B \lesssim 1.
\end{equation}

In the following derivation of this subsection, the Boyer-Lindquist
coordinates are used. Assume all the field lines threading the BH are
connected with the disk, then from the conservation of magnetic flux we have

\begin{equation}
\label{eq3} \Psi = \int {B_H (\rho \varpi )_H d\theta d\phi } = \int
{B_z \left( {\frac{\rho \varpi }{\sqrt \Delta }} \right)_D dr d\phi
} ,
\end{equation}

\noindent
where the subscripts ``H'' and ``D'' are used to indicate the quantities on
the horizon and the equatorial plane of the disk ($\theta = \pi / 2$),
respectively. The Boyer-Lindquist coordinates are given as

\begin{equation}
\label{eq4}
\begin{array}{l}
 \Sigma ^2 = ( {r^2 + a_\ast ^2 M^2} )^2 - a_* ^2 M^2\Delta
\sin ^2\theta ,\quad \rho ^2 = r^2 + a_\ast ^2 M^2\cos ^2\theta , \\
 \Delta = r^2 + a_\ast ^2 M^2 - 2Mr,\quad \varpi = (\Sigma / \rho ) \sin
\theta . \\
 \end{array}
\end{equation}

\noindent Since $\Delta=0$ at $r=r_H$, the lower boundary of the
integration interval in the second equality is set to be $r_H+\delta
r$, where $\delta r$ is a small quantity and taken as $\delta
r=0.01$\footnote{The influence of $\delta r$ can be ascribed to
$c_B$ as the effects of the MC process are mainly determined by the
strength of the field in the region $r>r_p$ (we will show this in
Sec.4).}. Substituting equation (\ref{eq1}) into equation
(\ref{eq3}) we get

\begin{equation}
\label{eq5} B_0 = \frac{\int {B_H (\rho \varpi )_H d\theta d\phi }
}{\int {F(r)\left( {\frac{\rho \varpi }{\sqrt \Delta }} \right)_D
drd\phi } } = \frac{2Mr_H \left( {1 - \cos \theta _0 } \right)B_H
}{\int {F(r)\left( {\frac{\rho \varpi }{\sqrt \Delta }} \right)_D dr
d\phi } } = 2 r_H k(a_\ast ,n) B_H / M.
\end{equation}

Given the configuration of the field, we can derive the energy and
angular momentum flux in the MC process by using the modified
equivalent circuit approach \citep{Wang02}. Considering a loop corresponds to
two adjacent flux surfaces (characterized by the magnetic flux $\Psi
$ and $\Psi + \Delta \Psi )$, the electromotive force due to the
rotation of the BH and the disk are expressed as

\begin{equation}
\label{eq6} \Delta \varepsilon _H = \left( {\Delta \Psi / 2\pi }
\right)\Omega _H ,\quad \Delta \varepsilon _D = - \left( {\Delta
\Psi / 2\pi } \right)\Omega ,\quad \Delta \Psi = 2\pi (\varpi \rho
)_H \Delta \theta \cdot B_H .
\end{equation}

\noindent The minus sign in the expression of $\Delta \varepsilon _D
$ arise from the direction of the flux. The parameter $\Omega $ is
the angular velocity of the ADAF, $\Omega _H = a_* / (2r_H )$ is the
angular velocity of the BH horizon.

The equivalent surface resistivity of the BH horizon is $4\pi$
\citep{MT82,TPM86}, while the surface resistivity of the disk is
$\sim 1 / (H\sigma ) = 4\pi \eta / H$, where $\eta \equiv 1 / (4\pi
\sigma )$ is the diffusivity of the magnetic field. As in many
papers \citep[e.g.][]{Lubow94,Lovelace95,Soria97}, we assume $\eta $
to be of the same order as the Shakura-Sunyaev (1973) kinematic $
\alpha$-viscosity coefficient, i.e., $\eta \sim \nu = \alpha c_s H$.
The resistances of the annular ring on the horizon and the disk are
thus
\begin{equation}
\label{eq7}
\Delta Z_H = 4\pi \cdot \frac{\rho _H \cdot \Delta \theta }{2\pi \varpi _H }
= \frac{2\rho _H \cdot \Delta \theta }{\varpi _H },
\end{equation}

\begin{equation}
\label{eq8} \Delta Z_D = \frac{1 }{H \sigma} \cdot \frac{\Delta
r}{2\pi \varpi _D } = \frac{2\alpha c_s \cdot \Delta r}{\varpi _D }.
\end{equation}

\noindent Thus the current in the loop is
\begin{equation}
\begin{array}{l}
\label{eq9}
 I = \frac{\Delta \varepsilon _H + \Delta \varepsilon _D
}{\Delta Z_H + \Delta Z_D } = \left( {\frac{\Delta \Psi }{2\pi }}
\right)\frac{\Omega _H - \Omega }{\Delta Z_H \cdot (1 + \xi )} =
\frac{1}{1 + \xi } \cdot \frac{a_\ast (1 - \beta _{HD} )}{2\csc
^2\theta - 1 + \sqrt {1 - a_\ast ^2 }
} \cdot MB_H . \\
(\xi \equiv \frac{\Delta Z_D }{\Delta Z_H } = \frac{\alpha c_s \varpi _H}{\rho_H \varpi_D } \left|\frac{\Delta r}{\Delta \theta
 }\right|;\quad \quad
\beta _{HD} \equiv \frac{\Omega }{\Omega _H })
\end{array}
\end{equation}

\noindent  In order to obtain $\xi$ and $I$ we have to find the
value of $\Delta r/\Delta \theta$ , which is related to the mapping
relation between the angular coordinate on the horizon and the
radial coordinate on the disk, i.e. $\theta(r)$.  According to the
conservation of the magnetic flux between the adjacent two flux
surfaces,
\begin{equation}
\label{eq10} d\Psi = B_H \cdot 2\pi \left( {\varpi \rho } \right)_H
d\theta = - B_z \cdot 2\pi \left( {\varpi \rho / \sqrt \Delta }
\right)_D dr.
\end{equation}
Substitute equations (\ref{eq1}) and (\ref{eq5}) into
the above equation, we get
\begin{equation}
\label{eq11} M\frac{d\cos \theta }{dr} = k(a_* ,n)\frac{\sqrt {r^4 +
M^2a_* ^2 r^2 + 2M^3a_* ^2 r} }{M\sqrt {r^2 + a_* ^2 M^2 - 2Mr}
}F(r) \equiv G(a_* ,r,n).
\end{equation}
\noindent Integrate equation (\ref{eq11}) and we obtain the mapping
relation:
\begin{equation}
\label{eq12} \cos \theta = \cos \theta _0 + \int_{r_H }^r {G(a_\ast
,r,n) \cdot \mbox{ }} dr.
\end{equation}

\noindent Our calculations show that, for the ADAF, the ratio of the
height to radius around $r_H$ is $\lesssim 0.3$, thus we assume
$\theta_0=0.4\pi$ so that $cot\theta_0 \approx 0.3$. Substitute
equation (\ref{eq11}) into equation (\ref{eq9}) we have
\begin{equation}
\label{eq13} \xi = \frac{2\alpha c_s G(a_\ast ,r,n)^{-1}}{\sqrt {r^2
+ a_\ast ^2 M^2 + 2M^3a_\ast ^2 r^{ - 1}} \left[ {2\csc ^2\theta - 1
+ \sqrt {1 - a_\ast ^2 } } \right]}
\end{equation}

 Since the current $I$ on the BH horizon feels Ampere's force, the
BH exerts a net torque on the magnetic flux tube
\begin{equation}
\label{eq14} \Delta T_{MC} = \varpi B_H I\rho \Delta \theta = \left(
{\frac{\Delta \Psi }{2\pi }} \right)I = \frac{4a_\ast (1 - \beta
_{HD} )G(a_\ast ,r,n)r_H }{\left( {1 + \xi } \right)\left( {2\csc
^2\theta - 1 + \sqrt {1 - a_\ast ^2 } } \right)} \cdot MB_H^2 \cdot
\Delta r.
\end{equation}
From the second equality it is easy to find that this torque
equals to the torque exerted on the disk by the same flux tube, or
equivalently speaking, the angular momentum flows between the BH and
the disk. The angular momentum flux can be written as
\begin{equation}
\label{eq15} H_{MC} = \frac{1}{4\pi r}\frac{\Delta T_{MC}
}{\Delta r}.
\end{equation}

\noindent The power transmitted to the disk through the tube is
given by

\begin{equation}
\label{pmc} \Delta P_{MC}=I\Delta \varepsilon_D+I^2\Delta
Z_D=4\pi r H_{MC}\Omega \Delta r+4\pi r H_{MC}(\Omega_F-\Omega)\Delta r
\equiv \Delta P_{MW}+\Delta Q_{Ohm},
\end{equation}

\noindent where
\begin{equation}
\label{omf} \Omega_F=\frac{\Omega_H\Delta Z_D+\Omega \Delta
Z_H}{\Delta Z_H+\Delta Z_D},
\end{equation}
\noindent is the angular velocity of the magnetic filed lines,
$\Delta P_{MW}\equiv 4\pi r H_{MC}\Omega \Delta r$ is the rate of
the mechanical work done by the electromagnetic torque on the disk
and $\Delta Q_{Ohm}\equiv 4\pi r H_{MC}(\Omega_F-\Omega)\Delta r$ is
the rate of Ohmic heating in the disk.

It is easy to calculate the power dissipated on the BH's stretched horizon
intersecting with the flux tube:

 \begin{equation}
\label{qbh} \Delta Q_{BH}=I^2\Delta Z_H,
 \end{equation}

\noindent which increases the irreducible mass of the BH.

\section{Basic Equations of the Accretion Flow}

We assume the energy and angular momentum transferred by the MC
process deposit into the accretion flow homogeneously in the
vertical direction. The height-averaged basic equations describing
the MCADAF can be written as
\begin{equation}
\label{eq18}
\dot {M} = - 4\pi r\rho H\upsilon = const,
\end{equation}
\begin{equation}
\label{eq19}
\upsilon \frac{d\upsilon }{dr} = (\Omega ^2 - \Omega _K^2 )r - \frac{1}{\rho
}\frac{dp}{dr},
\end{equation}
\begin{equation}
\label{eq20} \dot {M}\frac{d}{dr}(\Omega r^2) + 4\pi rH_{MC} = -
\frac{d}{dr}\left( {4\pi r^2\tau _{r\varphi } H} \right),
\end{equation}
\begin{equation}
\label{eq21} \rho \upsilon T_i \frac{ds_i }{dr} = (1 - \delta
)(q_{vis}^ + + \frac{Q_{Ohm} }{2H}) - q_{ie},
\end{equation}
\begin{equation}
\label{eq22} \rho \upsilon T_e \frac{ds_e }{dr} = \delta (q_{vis}^ +
+ \frac{Q_{Ohm} }{2H}) + q_{ie} - q^ -.
\end{equation}
Here $\dot {M}$ is the accretion rate, $c_s \equiv \sqrt{p/\rho}$ is
the isothermal sound speed, $p = p_{gas} / \beta _t = \rho c_s^2 /
\beta _t $ is the total pressure of the tangled magnetic field and
the gas pressure, $\beta _t $ is the ratio of the gas pressure to
the total pressure and is fixed at its ``typical'' value $\beta _t =
0.9$, $T$ is the temperature, $s$ is the entropy. The subscripts
``i" and ``e" indicate the quantities for ions and electrons,
respectively. The quantity $\tau _{r\varphi } = - \alpha p$ is the
$r\varphi $ component of the viscous stress tensor adopting the
$\alpha $ prescription \citep{SS73}, $H = c_s / \Omega _K $ is the
vertical scale height, $\Omega _K $ is the Keplerian angular
velocity calculated by using the pseudo-Newton potential given by
\citet{Muk02}, $\delta $ describes the fraction of the total energy
that directly heats the electrons and is set to be $\delta = 0.3$
following the detailed modeling result to the supermassive black
hole in our Galactic center \citep{YQN03}, $q_{vis}^ + = r\tau
_{r\varphi } \left(d\Omega/dr \right)$ is the heating rate of the
viscosity, $q_{ie} $ represents the volume energy transfer rate from
ions to electrons via Coulomb collisions, $q^ - $ is the cooling
rate of the electrons, which consists of bremsstrahlung,
synchrotron, and Comptonization \citep{NY95,Manmoto97}, and
$Q_{Ohm}=\Delta Q_{Ohm}/4\pi r \Delta r=H_{MC}(\Omega_F-\Omega)$ is
the rate of Ohmic dissipation per unit area of the disk.

Adopting the no-torque boundary condition at the horizon, we
integrate equation (\ref{eq20}) from $r_H$ to $r$ and get the
conservation equation for the angular momentum
\begin{equation}
\label{eq23}  l + \frac{\alpha rc_s^2 }{\upsilon } + \frac{1}{\dot
{M}}T^*_{MC} (r) = N_0 ,
\end{equation}
\noindent where
\begin{equation}
 \label{eq24}
T^*_{MC}(r)\equiv \int^r_{r_{out}}\frac{\partial T_{MC} }{\partial r}dr,
\end{equation}
\begin{equation}
\label{eq25} N_0 = l_0 + \frac{1}{\dot {M}}T^*_{MC}(r_H)=const ,
\end{equation}
with $l_0 $ being the angular momentum per unit mass
swallowed by the BH. The three terms on left-hand side of equation
(\ref{eq23}) correspond to the advected angular momentum,
viscous torque and magnetic torque due to the field lines in the
range from $r$ to $r_{out} $.

One more relation is given by the equation of state,
\begin{equation}
 \label{eq26}
 p_{gas}=k(T_i/\mu_i+T_e/\mu_e)/m_u,
\end{equation}
\noindent where $\mu$ is the mean molecular weight, $k$ is the
Boltzmann's constant, and $m_u$ is the atomic mass unit.

Thus we have a set of six equations including one integral, two
algebraic, and three differential equations, i.e. equations
(\ref{eq18}), (\ref{eq19}), (\ref{eq21})-(\ref{eq23}), and
(\ref{eq26}) for six unknown quantities, $H$ (or $c_s$), $v$,
$\rho$, $\Omega$, $T_i$, and $T_e$. This set of equations can be
solved with three outer boundary conditions and some given
parameters (we will specify them later). However, there is some
difficulty in obtaining $\Omega$ from the integral equation, viz.
equation (\ref{eq23}). So we use the first-order approximation, $
 T^*_{MC} (r) \approx T^*_{MC}(r + \Delta r)
- \left.\partial T_{MC} /\partial r \right|_{r + \Delta r} \Delta r.
 $

\section{Numerical Results}

We adopt $M = 10M_ \odot$ and $r_{out} = 10^3M$ in this paper.
Regarding the outer boundary of the ADAF, the ion temperature $T_i$
should be of the same order as the virial temperature, and the
electron temperature $T_e$ should be somewhat lower than $T_i$
because of the radiation of the electrons. The angular velocity of
the accreting flow should be sub-Keplerian \citep{NMQ98}. So we
impose the boundary conditions as $T_i = 2\times 10^9\mbox{K}$, $T_e
= 1\times 10^9\mbox{K}$, $\upsilon / c_s = 0.3$. At last, by
adjusting the eigenvalue of the problem, $N_0$, we can obtain the
global transonic solution, i.e., a solution that can pass through
the sonic point smoothly.

The free parameters of our MCADAF model include $a_*$, $c_B $,
$\lambda $, $n$, $\alpha $, and $\dot {m}$, where $\dot {m} = \dot
{M} / \dot {M}_{\rm Edd} $ with the Eddington accretion rate $\dot
{M}_{\rm Edd} = 1.39\times 10^{18}(M/M_ \odot) \mbox{g} \cdot
\mbox{s}^{ - 1}$. The first parameter describes the spin of the BH,
the next three are associated with the magnetic field, and the last
two describe the ADAF.

\begin{figure}
\epsscale{0.9}\plotone{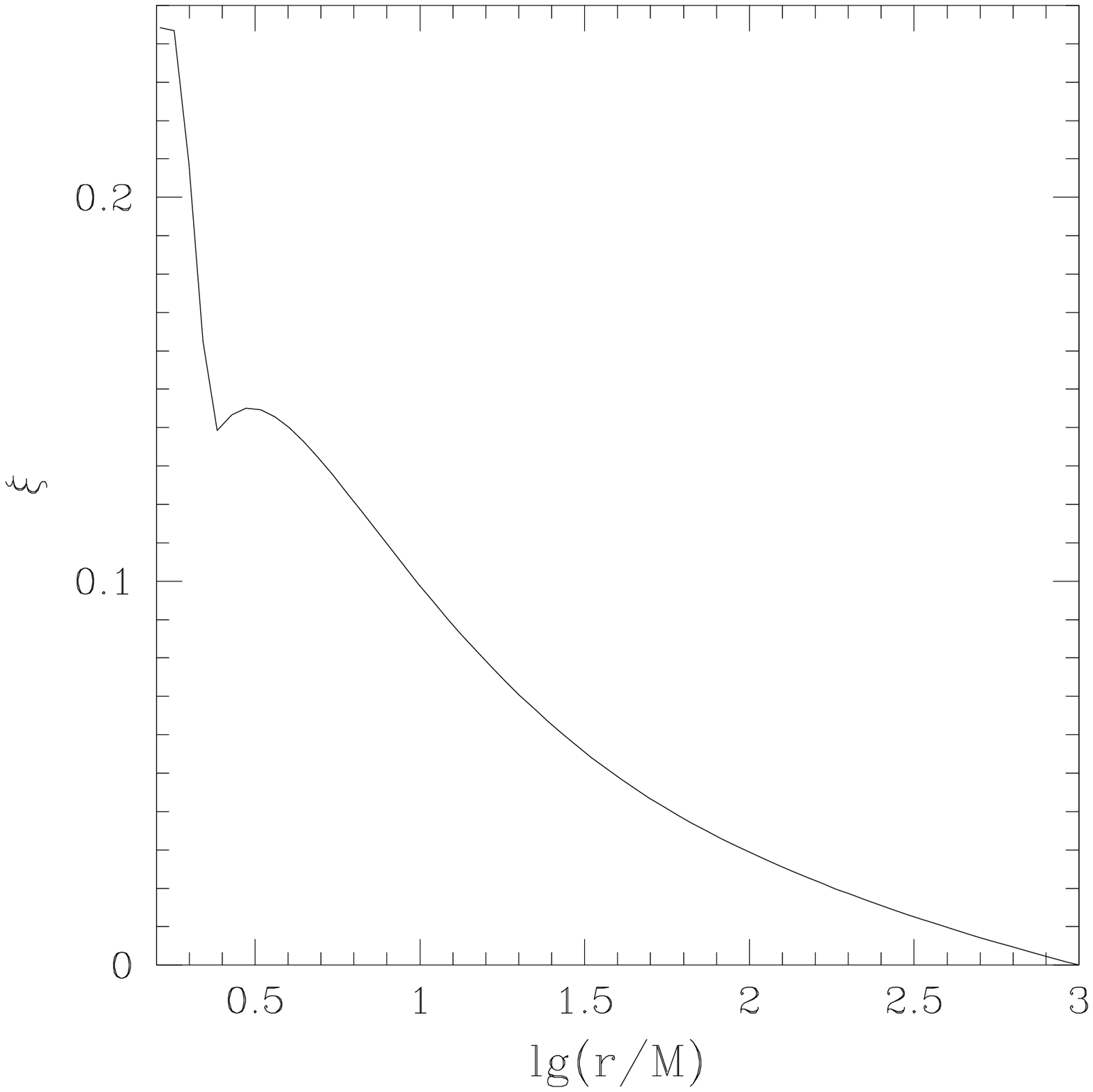} \caption{\label{fzita} Profile of
$\xi(\equiv\Delta Z_D/\Delta Z_H)$ on the disk. The parameters are
$a_*=0.9, n=3, \lambda=1$ (corresponding to
$r_p=r_{ms}$), $c_B = 1$, $\alpha=0.3$ and $\dot{m}=0.01$.}
\end{figure}

\begin{figure}
\epsscale{0.9}\plotone{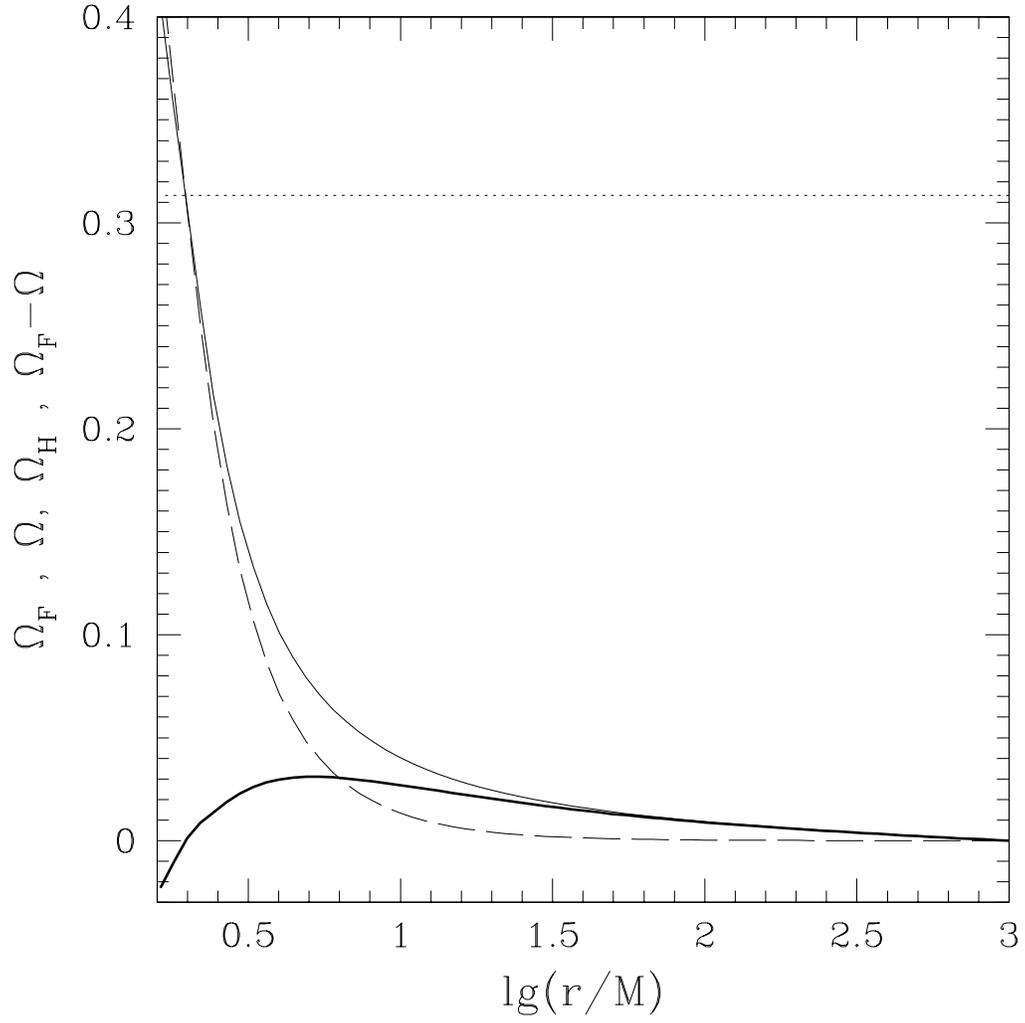} \caption{\label{fomeg} Curves of the
angular velocities. The solid, dashed and dotted lines correspond to
$\Omega_F$, $\Omega$ and $\Omega_H$, respectively. The thick solid
line shows $\Omega_F-\Omega$. The parameters are the same as Figure 1.}
\end{figure}

Figure \ref{fzita} shows the curve of $\xi(\equiv \Delta Z_D/\Delta
Z_H)$ when $a_*=0.9, n=3, \lambda=1$ (corresponding to
$r_p=r_{ms}$), $c_B = 1$, $\alpha=0.3$ and $\dot{m}=0.01$. As the
figure shows, the resistance of the disk is small compared with the
resistance on the stretched horizon. Especially, at the outer
boundary, the resistance of the disk is completely negligible. Since
the distribution of the magnetic field we assumed is not smooth,
there is a break at $r_{ms}$. Figure \ref{fomeg} shows the curves of
$\Omega,\Omega_F$, $\Omega_H$ and $\Omega_F-\Omega$ for the same
parameters as Figure \ref{fzita}. From this figure we find that
$\Omega_F$ always lies between $\Omega_H$ and $\Omega$, which agrees
with equation (\ref{omf}). We also find that the relative angular
velocity of the magnetic field lines to the disk, i.e.
$\Omega_F-\Omega$, achieves maximum at some radius between the inner
and outer boundaries. This is a natural result because
$\Omega_F-\Omega$ is proportional to the product of two factors,
$\Delta Z_D/(\Delta Z_H+\Delta Z_D)$ and $\Omega_H-\Omega$ (see
equation (\ref{omf})), which approach to zero at the outer and inner
boundaries, respectively. At the innermost region, $\Omega
>\Omega_H$. This unphysical result arise because we do not use the
exact general relativity.

\begin{figure}
\epsscale{0.9}\plotone{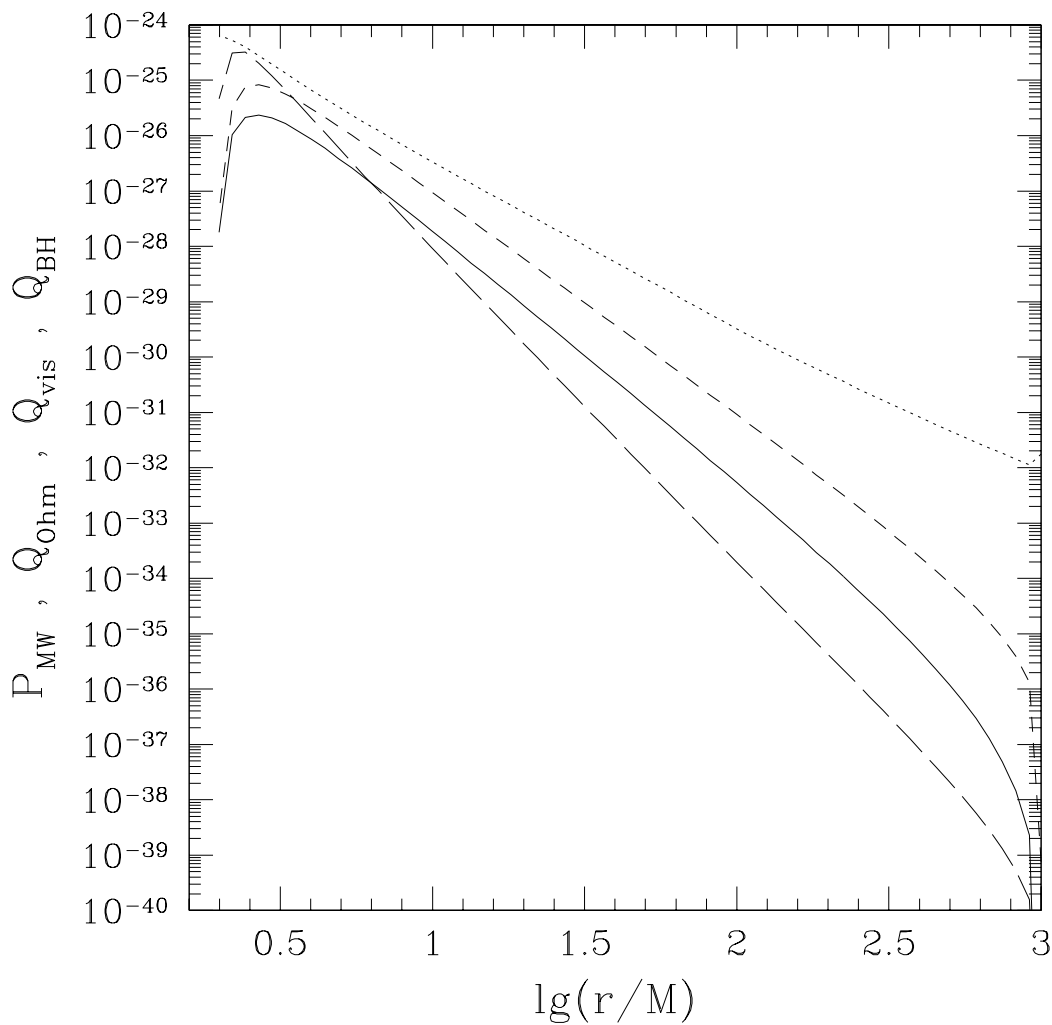}\caption{\label{fq} Curves of the rate
of mechanical work due to electromagnetic torque and the heating
rates per unit area. The solid line represents the rate of Ohmic
heating in the disk $Q_{Ohm}$, the long dashed line shows the power
of the electromagnetic torque on the disk $P_{MW}$, the short dashed
line shows the Ohmic dissipation on the BH's stretched horizon
$Q_{BH}$. As comparison the curve of
 viscous heating rate per unit area $Q_{vis}$ is shown in dotted line.
The parameters are the same as Figure 1.}
\end{figure}

From Figure \ref{fq} we can get some ideas of the partitioning of
the magnetically-extracted rotational energy of the BH. In this
figure we show the curves of the rate of mechanical work due to
electromagnetic torque and three kinds of heating rates per unit
area. The solid line represents $Q_{Ohm}$, the
 long dashed line for $P_{MW}=\Delta P_{MW}/4\pi r \Delta r=H_{MC}\Omega$,
the short dashed line for the Ohmic dissipation on the BH's
stretched horizon, $Q_{BH}=\Delta Q_{BH}/4\pi r \Delta r$, while the
dotted line for the viscous heating rate per unit area,
 $Q_{vis}=2Hq^+_{vis}$. As the figure shows, $Q_{Ohm}$ dominates over
$P_{MW}$ in the outer region of the disk while the latter dominates
in the inner region. Compare $Q_{Ohm}+P_{MW}$ with $Q_{BH}$, we find
the efficiency of extracting energy from the BH to the disk is very
small except in the inner region. Moreover, it can be seen that the
MC power is small in contrast to the viscous heating rate.

\begin{figure}
\epsscale{0.9}
 \plotone{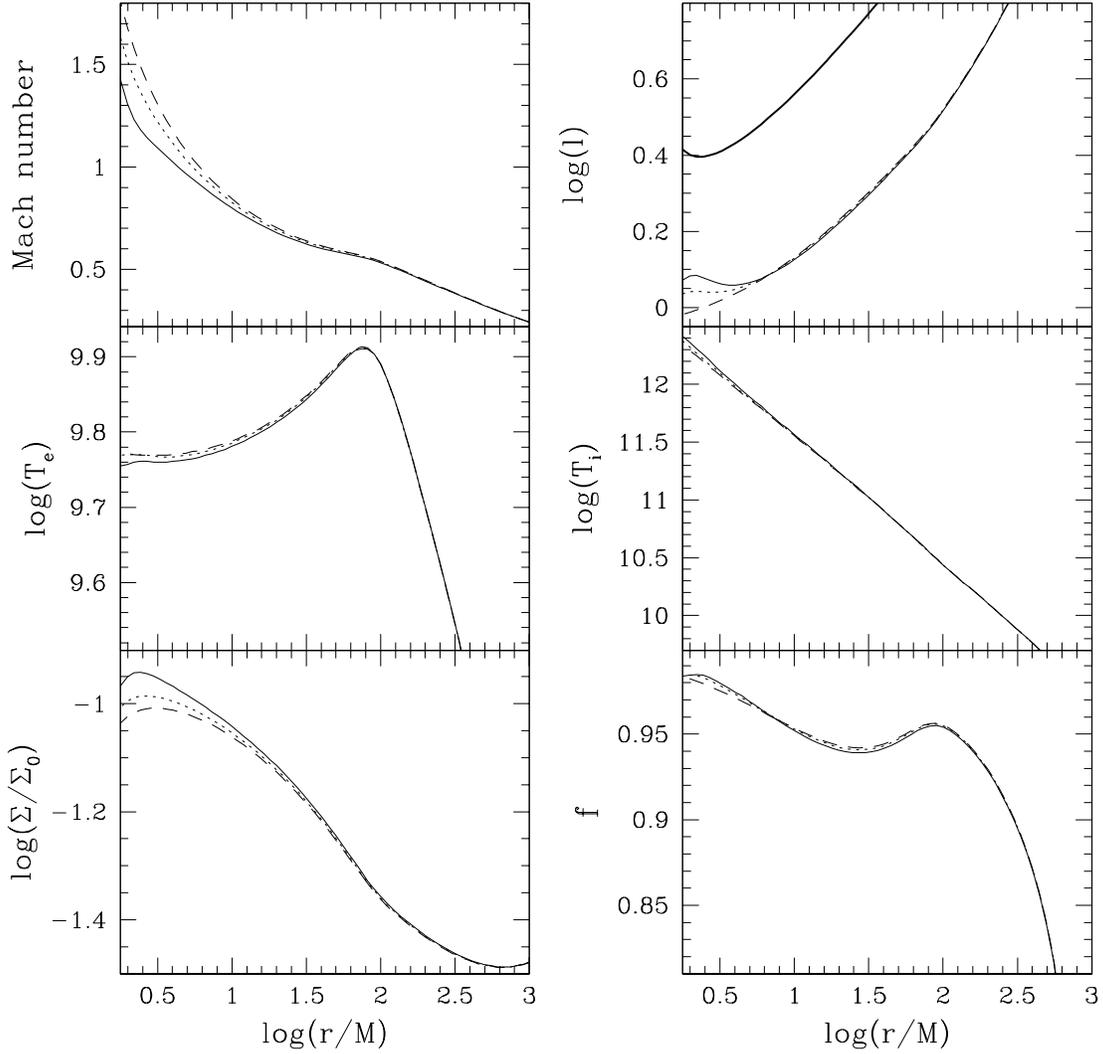}  \caption{\label{fk} The profiles of
the radial Mach number, specific angular momentum $l$, electron and
ion temperatures $T_e$ and $T_i$, surface density $\Sigma$ (in unit
of $\Sigma_0\equiv \dot{M}/M$), and the advection factor $f$ for
different $c_B $. Other parameters are the same as Figure 1. The
solid, dotted, and dashed lines correspond to $c_B = 1$, 0.5, and 0,
respectively. The thick solid line in the top right panel shows the
Keplerian angular momentum.}
\end{figure}

The profiles of the radial Mach number, surface density $\Sigma$,
$T_e $ and $T_i $, specific angular momentum $l$, and the advection
factor $f$ ($ =q_{adv}/(q_{vis}^++F_{MC}/2H)$)
 of the accretion flow for different $c_B $
are shown in Figure \ref{fk}. The solid, dotted, and dashed lines
correspond to $c_B = 1$, 0.5, and 0, respectively.

From Figure \ref{fk} we find that, when the MC process is present,
the sonic point moves inward, $T_e $ decreases, $\Sigma$ and $T_i $
increase, while $l$ and $f$ decrease in the outer region and
increase in the inner region. These effects can be understood as
follows. As the spin of the BH is very fast ($a_*=0.9$), angular
momentum and energy are transferred from the BH to the disk. The
energy flux raises the temperature of the ions. The angular momentum
flux hinders the infalling of the accreting material. Thus the sonic
point moves inward and the surface density increases. Since the
optical depth is proportional to the surface density, the Compton
cooling rate goes up, and consequently, the temperature of the
electrons decreases. Compared with the case without MC process, the
$r\varphi$ component of the viscous stress tensor in a MCADAF around
a fast rotating BH is a bit larger due to the higher pressure $p$ in
the outer region, so the angular momentum is transferred more
efficiently and the specific angular momentum there is smaller. But
at small radius, the increase of the angular momentum due to the MC
process dominates over the decrease due to the viscous torque, so
that the specific angular momentum can even increase, as can be seen
from equation (\ref{eq20}). Similarly, in the outer region of the
disk the radiative cooling rate becomes higher and $f$ decreases,
while in the inner region $f$ goes up because the heating rate due
to the MC process increases more quickly than the radiative cooling
rate does.  Additionally, as the magnetic field threading the BH
becomes stronger, the influences of the MC process become more
significant. However, as Figure \ref{fq} shows, the contribution of
the MC process is of less importance, so its influences are small.

\begin{figure}
\epsscale{0.9} \plotone{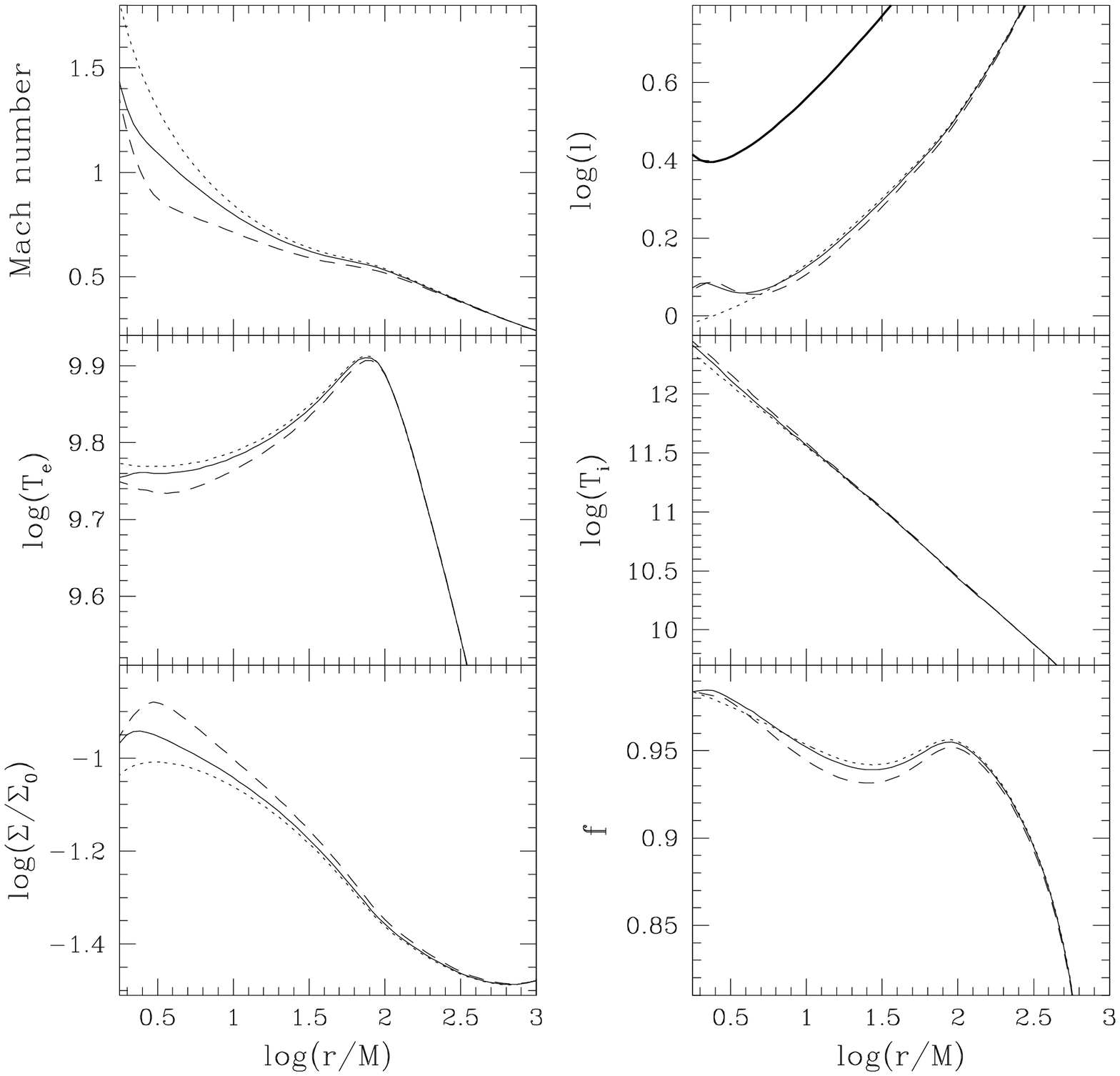}
  \caption{\label{fla} Same with Figure 1, but for
different $\lambda $ when $a_* = 0.9$, $c_B = 1$, $\alpha = 0.3$,
and $\dot {m} = 0.01$. The solid and dashed lines correspond to
$\lambda = 1$ and 1.5, respectively. The dotted line corresponds to
the case without MC process.}
\end{figure}

Figure \ref{fla} shows the influences of the parameter $\lambda$
(ref. the paragraph below eq. 1). The solid and dashed lines are for
$\lambda= 1$ and $1.5$, respectively. For comparison purpose the
dotted line is shown for the case without MC process. From this
figure we see that the effects of increasing $\lambda$ are similar
to those of increasing $c_B$. This is because the magnetic field in
the region $r>r_p$ strengthens as these two parameters increase.
Although the increase of $\lambda$ also leads to the decrease of the
magnetic field in the region $r_H<r<r_p$, the MC effects are small
in this region, because the gravitational force there is so strong
and the radial velocity is so high that the energy and angular
momentum transferred by the MC process do not play any significant
role.

\begin{figure}
\epsscale{0.9} \plotone{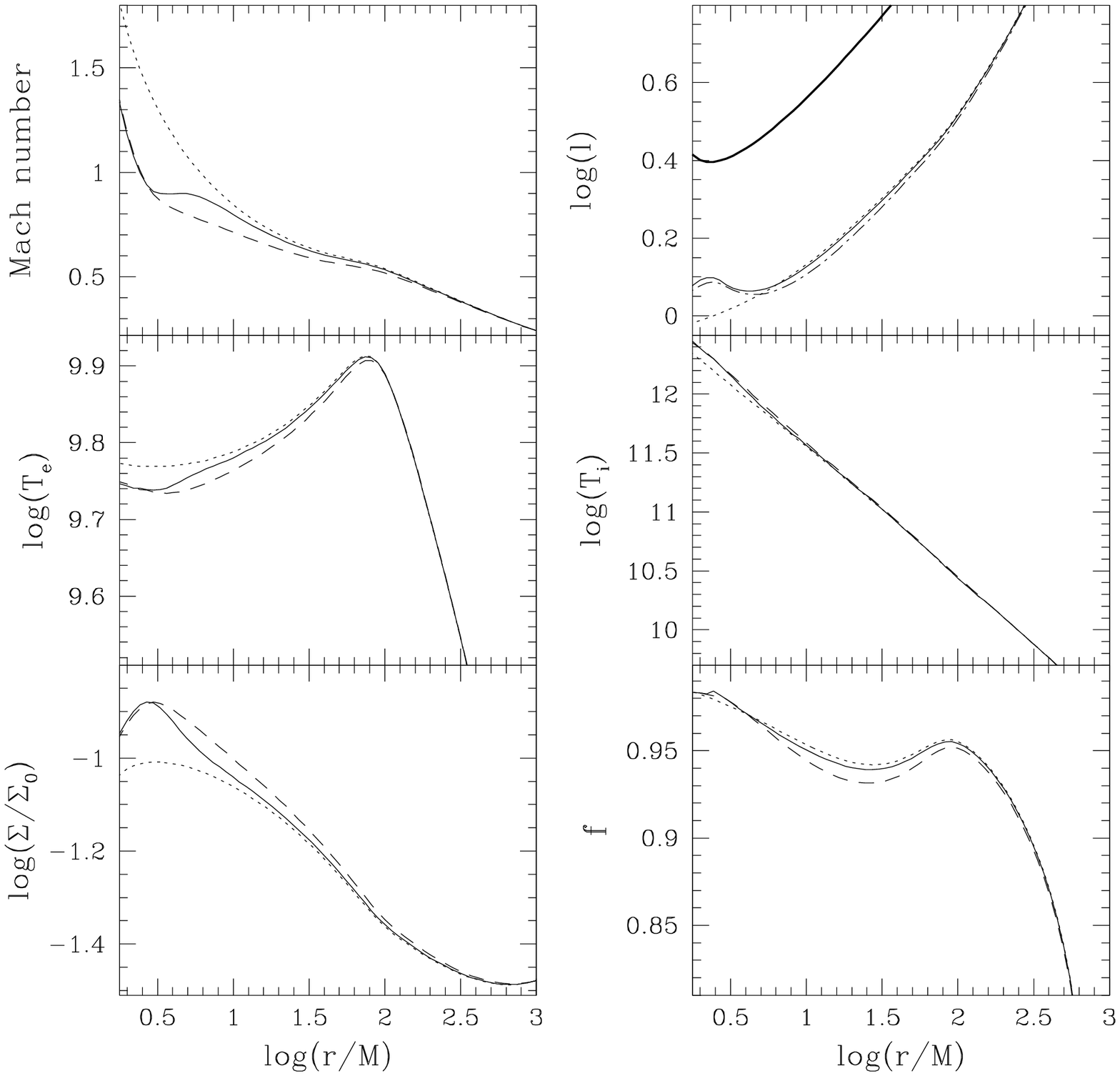}
 \caption{\label{fn} Same with Figure 1, but for
different $n$ when $a_* = 0.9$, $c_B = 1$, $\alpha = 0.3,
\lambda=1.5$, and $\dot {m} = 0.01$. The solid and dashed lines
correspond to $n = 4$ and 3, respectively. The dotted line
corresponds to the case without MC process.}
\end{figure}

Figure \ref{fn} shows the effects of parameter $n$ (ref. eq. 1). The
solid and dashed lines are for $n=4$ and $3$, respectively. The
lines for the case without MC process are shown with the dotted
lines. From this figure we find that the MC effects are more
significant for smaller $n$. It is because the magnetic field is
weaker in the outer region ($r > r_p$) when $n$ is bigger, which can
be seen from equations (\ref{eq1}) and (\ref{eq5}).

\begin{figure}
\epsscale{0.6} \plotone{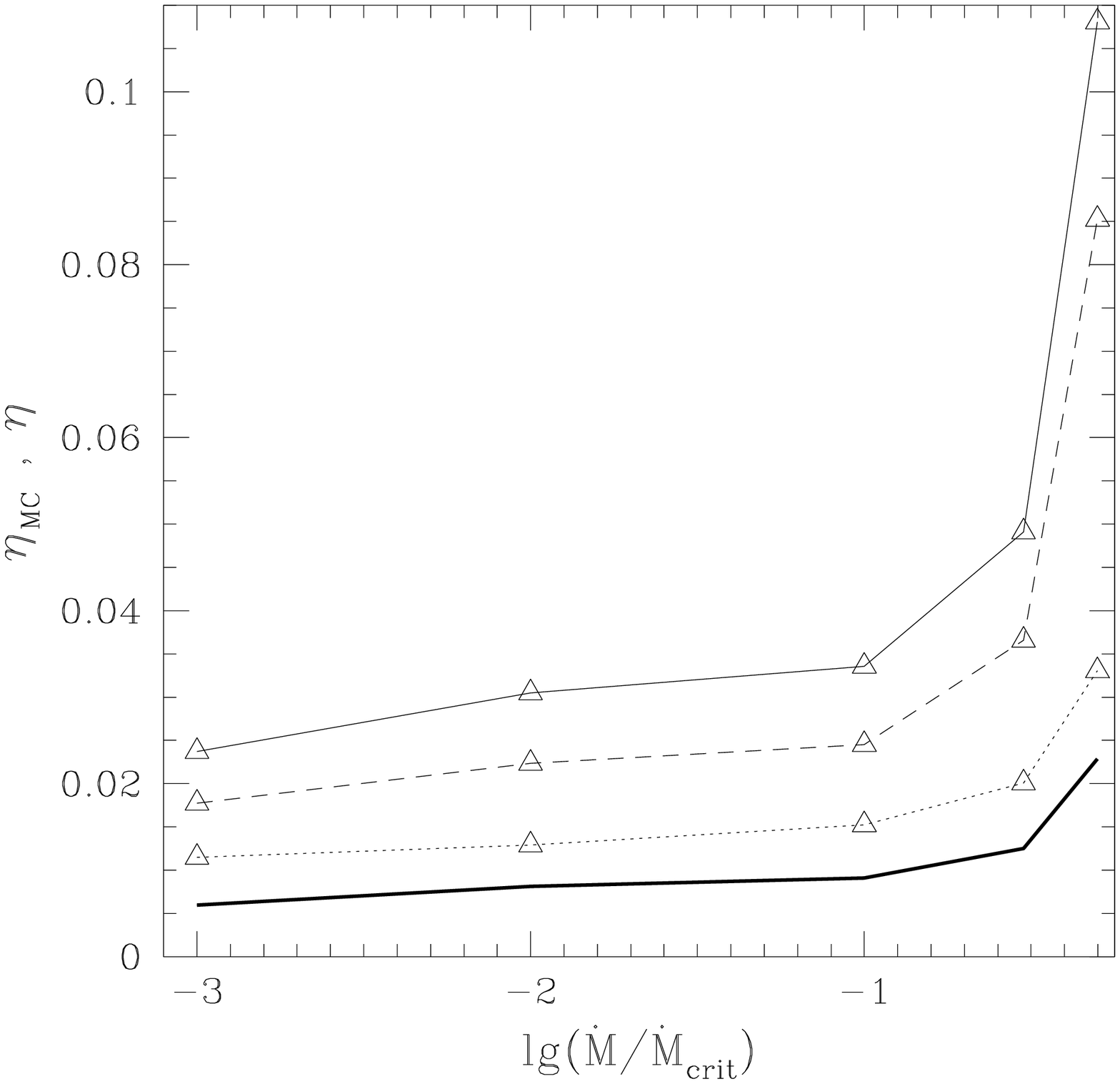} \\ \plotone{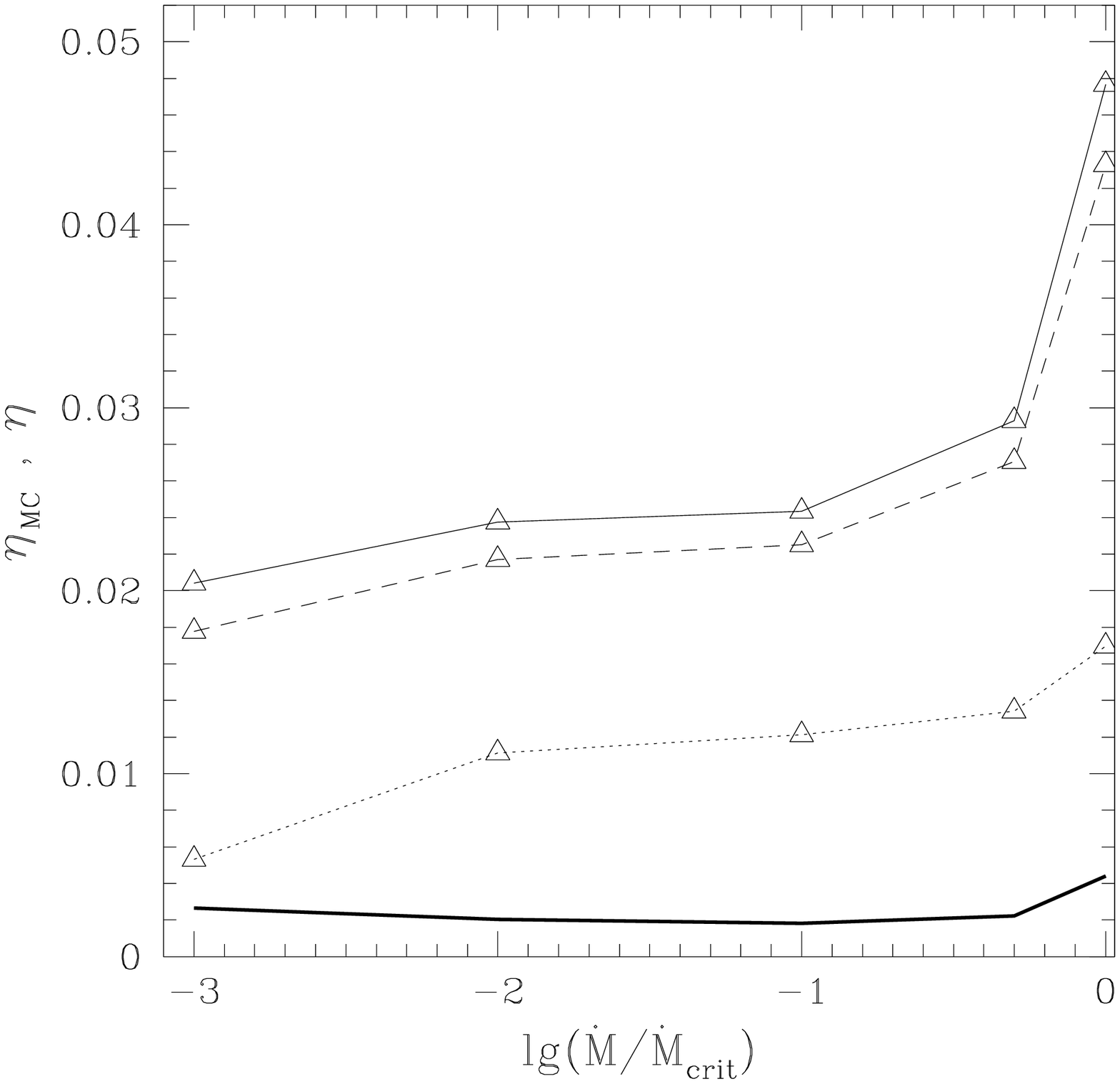}

 \caption{\label{feta} The radiative efficiency of the accretion flow
as a function of the accretion rate. Here $\dot{M}_{\rm crit}\equiv
\alpha^2\dot{M}_{\rm Edd}$. The thin solid lines correspond to the
results of the present MCADAF model ($a_*=0.9$, $c_B=1.0$, $\lambda
=1.0$ and $n=3$), the dashed lines indicate the results without MC
process, the thick solid lines show the efficiency improved by the
MC process ($\eta_{MC}$), and the dotted lines show the results of a
Schwarzschild BH without considering the MC process. In the upper
panel $\alpha=0.3$ and in the lower panel $\alpha=0.1$.}
\end{figure}

We calculate the influence of the MC process on the radiative
efficiency of the ADAF. The results are shown in Figure \ref{feta}.
The parameters are $a_*=0.9, c_B=1.0, \lambda=1$ and
$n=3$. The quantity $\dot {M}_{crit}$ denotes the critical accretion
rate of an ADAF, which is $\sim \alpha ^2\dot {M}_{\rm Edd} $
\citep[e.g.][]{NMQ98}. The radiative efficiency increased by the MC
process is written as $\eta _{\rm MC} = \eta _{\rm MCADAF} - \eta
_{\rm ADAF} $, where $\eta _{\rm MCADAF} $ and $\eta _{\rm ADAF} $
are the efficiencies of the MCADAF and pure ADAF, respectively. From
Figure \ref{feta} we find that, when $\alpha=0.3$, the MC process
can raise the efficiency by about one percentage point, i.e., $\sim
30\% \eta_{\rm ADAF}$. But when $\alpha=0.1$ the effect of the MC
process is very weak. The efficiency goes up because of two reasons:
firstly, the MC process transports additional energy to the ADAF;
secondly the angular momentum transported by the MC process
decreases the radial velocity of the ADAF, and thus makes the ADAF
more efficient in radiating. When $\alpha $ is smaller, the specific
angular momentum of the accreting material is larger since viscosity
is less efficient in moving angular momentum out, and consequently,
the difference between the angular velocities of the BH and the disk
is smaller. Considering equations (\ref{eq14}), (\ref{eq15}) and
(\ref{pmc}), the angular momentum and energy transported by the MC
process decrease.

In all the above discussions, the spin of the BH is very large. The
angular momentum and energy are transferred from the BH to the disk.
If the BH rotates slowly, the angular momentum and energy may be
transferred from the disk to the BH. However, since $H_{MC} $ is
proportional to $a_*$, the MC effect is not so significant as the
case when $a_* = 0.9$. In addition, there is a critical value of
$a_*$, at which the total energy and angular momentum transmitted by
the MC process are zero and consequently $\eta_{MC}=0$. In the SSD
case this value is about $a_*=0.283$ for $n = 3$ \citep{Wang03},
whereas in our MCADAF model, it is about 0.172. The critical value
is smaller in the MCADAF case because the angular velocity of the
ADAF is sub-Keplerian.

 The spin of the BH can even be negative, i.e., retrograde spin.
From equations (\ref{eq14}) and (\ref{eq15}) it is easy to find that
the angular momentum always flows from the ADAF to the BH when
$a_*<0$. If the resistance of the ADAF is zero, i.e. $\xi=0$ or
$\Omega_F=\Omega$, the energy flows in the same direction as that of
the angular momentum, as can be seen from equation (\ref{pmc}). If
$\xi$ is nonzero and big enough, Ohmic dissipation in the disk may
offset the loss of energy that conveyed to the BH. According to
equation (\ref{pmc}), the critical condition is $\Delta P_{MC}=0$,
or equivalently, $\xi+\beta_{HD}=0$. Our calculations show that when
$a_*<0$, $\xi$ is so small that $\xi+\beta_{HD}<0$ holds for almost
all cases, and the net energy flows from the ADAF to the BH. If the
BH rotates rapidly, e.g. $a_*=-0.9$, the effects of the MC process
will be negative  compared to the case of $a_*=0.9$: the radial
velocity and the temperature of the electrons increase, the sonic
point moves outward, the temperatures of the ions decrease, etc.

\section{Summary and Discussion}

In this paper we investigate the influence of the magnetic coupling
process on the dynamics of the ADAF. The effects of the MC process
on the basic equations of the accretion flow is simplified as a
source of angular momentum and energy.  The angular momentum and
energy fluxes are derived with the equivalent circuit approach.
We find that when
the BH rotates fast (e.g., $a_*=0.9$) and when the viscous parameter
$\alpha$ is large, $\alpha=0.3$, for reasonable magnetic field, the
MC process can mildly affect the dynamics of the ADAF, increasing
the ion temperature and the density of the accretion flow and
decreasing the electron temperature and the radial velocity. The MC
process can also raise the effciency of the ADAF by  $\sim 30\%$.
But if $\alpha=0.1$ or smaller, the influences of the MC process are
much weaker and can be neglected.

In the above calculations the strength of the magnetic field is
estimated following \citet{MSL}, which is equivalent to assume $c_B
\le 1$. Obviously uncertainties exist in the above estimation. On
the one hand, recent MHD simulations show that the magnetic field
strength near the horizon can be very high, almost four times as
large as the the equipartition value \citep{McKinney05}, which
corresponding roughly to $c_B\approx 2$. On the other hand, our
calculation requires that there exists an upper limit to the value
of $c_B$. This is because if $c_B$ were too large, the transferred
angular momentum from the BH to the accretion flow would be so
significant that the accretion could not proceed due to the strong
centrifugal force. We find that the highest value of $c_B$ depends
on the accretion rate for given $a_*$. It can be $\sim 5$ if the
accretion rate is very low but $\sim 1$ if the accretion rate is as
high as $\dot{m}\ga 0.1$. Considering the above two limitations on
$c_B$, the increased efficiency due to the MC process $\eta_{\rm
MC}(\equiv \eta_{\rm MCADAF}-\eta_{\rm ADAF})$ can be as high as
$10\%$.

Observations of the hard state of BH X-ray binaries sometimes show
luminosities as high as $L_x \sim 10-30\% L_{\rm Edd}$, which are
much higher than $4\% L_{Edd}$, the highest luminosity that an ADAF
surrounding a Schwarzschild BH can produce \citep{E97}. When the
accretion rate is higher than the critical rate of an ADAF, the
accretion flow enters into the regime of the Luminous Hot Accretion
Flow (LHAF) \citep{Yuan01}. \citet{Yuanetal07} found that the
highest luminosity an LHAF surrounding a non-rotating BH (with
$\dot{M}=0.3\dot{M}_{\rm Edd}$) can produce was $\sim 8{\%}L_{\rm
Edd}$, which is still too low. They speculated that when the spin of
the BH and the MC process were taken into account, the highest
luminosity an LHAF could produce would possibly be high enough to
explain the observed high $L_{x}$. Now, taking into account the black
hole spin and the MC process, we recalculate the maximum luminosity
an LHAF (with $\dot{m}=0.3$) can produce, with reasonable parameters
such as $a_*=0.9$, $\lambda=1, c_B\approx 1$ (note this is the
largest possible value when $\dot{m}=0.3$) and $n=3$. We find the highest luminosity is
$\sim 14\% L_{\rm Edd}$. The increase due to the MC process is $\sim
1.0 \% L_{\rm Edd}$ while that due to the BH spin is $\sim 5.1\%
L_{\rm Edd}$. So the MC process seems not so helpful to increase the
highest luminosity an LHAF can produce to explain the observed
highest $L_x$.

 In this paper we assume that the closed field extends to the
outer boundary of the ADAF according to a power law. As a matter of
fact the magnetic connection between the BH and the disk can be
maintained only within a limited radius. \cite*{Wang04} discussed
the constraint of the screw instability to the MC region of a SSD
based on the Kruskal-Shafranov criterion: the screw instability will
occur, if the magnetic field line turns around itself about once
\citep{K66,B78}. It turns out that the MC region is limited within
some critical radii on the SSD. By numerically solving the
Grad-Shafranov equation, the main differential equation that
describes the structure of the magnetosphere, \cite*{U05} argued
that for a rapidly-rotating BH the field lines are frame-dragged by
the BH so much, and the toroidal magnetic field becomes so strong
that the magnetic connection between the BH and the disk cannot be
maintained over a large range of radii on the disk. We shall address
this issue in the context of ADAF in our further work.

\acknowledgments We thank the anonymous referee for his/her very
constructive suggestions and careful examination, which greatly
improve the presentation. This work was supported by the
One-Hundred-Talent Program of CAS and the Pujiang Program.

\clearpage

\begin{thebibliography}{}
\bibitem[Abramowicz et al.(1995)]{A95} Abramowicz, M. A., Chen, X.,
Kato, S., Lasota, J.-P., \& Regev, O. 1995, \apj, 438, L37
\bibitem[Bateman(1978)]{B78} Bateman, G. 1978, in MHD Instabilities
(Cambridge: MIT)
\bibitem[Blandford \& Znajek(1977)]{BZ77} Blandford, R. D., \& Znajek, R. L. 1977, \mnras, 179, 433
\bibitem[Blandford(1999)]{Blandford99} Blandford, R. D. 1999, in Astrophysical Discs, ed. J. A. Sellwood, {\&}
J.Goodman, ASP Conf. Ser., 160, 265 (astro-ph/9902001)
\bibitem[De Villiers et al.(2003)]{DeV03} De Villiers, J., Hawley,
J. F., \& Krolik J. H. 2003, \apj, 599, 1238
\bibitem[Esin et al.(1997)]{E97} Esin, A. A., McClintock, J. E., \&
Narayan, R. 1997, \apj, 489, 865
\bibitem[Hawley(2000)]{Hawley00} Hawley, J. F. 2000, \apj, 528, 462
\bibitem[Hawley \& Balbus(2002)]{HB02} Hawley, J. F., \& Balbus, S.
A. 2002, \apj, 573, 738
\bibitem[Hirose et al.(2004)]{Hirose04} Hirose, S., Krolik, J. H., De
Villiers, J., \& Hawley, J. F. 2004, \apj, 606, 1083
\bibitem[Kato, Fukue \& Mineshige(1998)]{KFM98} Kato, S., Fukue, J., \&
Mineshige, S., eds. 1998, Black-Hole Accretion Disks, (Tyoto: Kyoto
Univ. Press)
\bibitem[Kadomtsev(1966)]{K66} Kadomtsev, B. B. 1966, Rev. Plasma
Phys., 2, 153
\bibitem[Koide(2003)]{Koide03} Koide, S. 2003, \prd, 67, 104010
\bibitem[Lai(1998)]{Lai98} Lai, D. 1998, \apj, 502, 721
\bibitem[Lee(1999a)]{Lee99a} Lee, U. 1999a, \apj, 511, 359
\bibitem[Lee(b)]{Lee99b} Lee, U. 1999b, \apj, 525, 386
\bibitem[Li \& Pacz\'ynski(2000)]{Li00} Li, L. X., {\&} Paczynski, B. 2000, \apj, 534, L197
\bibitem[Li(2002)]{Li02} Li, L. X. 2002, \apj, 567, 463
\bibitem[Livio et al.(1999)]{Livio99} Livio, M., Ogilvie, G. I., \&
Pringle, J. E. 1999, \apj, 512, 100
\bibitem[Lovelace et al.(1995)]{Lovelace95} Lovelace, R.V.E., Romanova, M.M., \& Bisnovatyi-Kogan, G. S. 1995 \mnras,
275, 244
\bibitem[Lubow et al.(1994)]{Lubow94} Lubow, S. H., Papaloizou, J. C. B., \& Pringle, J. E., 1994, \mnras, 267, 235
\bibitem[Macdonald \& Thorne(1982)]{MT82} Macdonald, D., \& Thorne, K. S. 1982, \mnras, 198, 345
\bibitem[Manmoto et al.(1997)]{Manmoto97} Manmoto, T., Mineshige, S., \& Kusunose, M. 1997, \apj, 489, 791
\bibitem[McKinney(2005)]{McKinney05} McKinney, J. C. 2005, \apj, 630, L5
\bibitem[McKinney \& Gammie(2004)]{MG04} McKinney, J. C., \& Gammie,
C. F. 2004, \apj, 611, 977
\bibitem[Moderski, Sikora \& Lasota(1997)]{MSL} Moderski, R., Sikora, M., \& Lasota, J. P. 1997, in Proceedings
of the International Conf., Relativistic Jets in AGNs, ed. M.
Ostrowski, M. Sikora, G. Madejski, \& M. Begelman (astro-ph/9706263)
\bibitem[Mukhopadhyay(2002)]{Muk02} Mukhopadhyay, B. 2002, \apj, 581, 427
\bibitem[Narayan(2005)]{Narayan05} Narayan, R. 2005, ApS\&S, 300, 177
\bibitem[Narayan \& Yi(1994)]{NY94} Narayan, R., \& Yi, I., 1994, \apj, 428, L13
\bibitem[Narayan \& Yi(1995)]{NY95} Narayan, R., \& Yi, I. 1995, \apj, 452, 710
\bibitem[Narayan, Mahadevan \& Quataert(1998)]{NMQ98} Narayan, R., Mahadevan, R., \&
Quataert, E. 1998, in "The Theory of Black Hole Accretion Discs",
eds. M. A. Abramowicz, G. Bjornsson, and J. E. Pringle, (Cambridge
University Press)
\bibitem[Novikov \& Thorne(1973)]{NT73} Novikov, I. D., \& Thorne,
K. S., in Black Holes, ed. C. DeWitt and B. DeWitt (New York: Gordon
\& Breach), 345
\bibitem[Page \& Thorne(1974)]{PT74} Page, D. N., \& Thorne, K. S.,
1974, \apj, 191, 499
\bibitem[Shakura \& Sunyaev(1973)]{SS73} Shakura, N. I., Sunyaev, R. A. 1973, \aap 24, 337
\bibitem[Soria et al.(1997)]{Soria97} Soria, R., Li, J., \& Wickramasinghe, D. T. 1997, \apj, 487, 769
\bibitem[Thorne et al.(1986)]{TPM86} Thorne, K. S., Price, R. H., \& Macdonald, D. A. 1986, in Black Holes: The Membrane
Paradigm, (New Haven and London: Yale Univ Press)
\bibitem[Tout \& Pringle(1996)]{Tout96} Tout, C. A., \& Pringle, J.
E. 1996, \mnras, 281, 219
\bibitem[Uzdensky(2004)]{U04} Uzdensky, D. A. 2004, \apj, 603, 652
\bibitem[Uzdensky(2005)]{U05} Uzdensky, D. A. 2005, \apj, 620, 889
\bibitem[Wang et al.(2002)]{Wang02} Wang, D. X., Xiao, K., {\&} Lei, W. H. 2002, \mnras,
335, 655
\bibitem[Wang et al.(2003)]{Wang03} Wang, D. X., Lei, W. H., \& Ma, R. Y. 2003, \mnras, 342, 851
\bibitem[Wang et al.(2004)]{Wang04} Wang, D. X., Ma, R. Y., Lei, W. H., \& Yao, G. Z. 2004, \apj, 601, 1031
\bibitem[Ye et al.(2007)]{Ye07} Ye, Y. C., Wang, D. X., \& Ma, R. Y.
2007, \na, 12, 471
\bibitem[Yuan(2001)]{Yuan01} Yuan, F. 2001, \mnras, 324, 119
\bibitem[Yuan(2007)]{Yuan07} Yuan, F. 2007, to appear in "The Central
Engine of Active Galactic Nuclei", ed. L. C. Ho and J.-M. Wang
(San Francisco: ASP) (astro-ph/0701638)
\bibitem[Yuan, Quataert \& Narayan(2003)]{YQN03} Yuan, F., Quataert, E., \& Narayan, R. 2003,
ApJ, 598, 301
\bibitem[Yuan et al.(2007)]{Yuanetal07} Yuan, F., Zdziarski, A. A., Xue,
Y. Q., \& Wu, X. B. 2007, \apj, 659, 541



\end{thebibliography}
\end{document}